\documentclass[twocolumn]{autart}    

\usepackage{balance}

\usepackage{hyperref}
\hypersetup{
    colorlinks,
    citecolor=blue,
    filecolor=blue,
    linkcolor=blue,
    urlcolor=blue
}

\usepackage[pdftex]{graphicx}

\usepackage{enumerate}

\usepackage{mathrsfs,amsmath}
\usepackage{amssymb}
\usepackage{hyperref}

\usepackage{natbib}

\newtheorem{theorem}{Theorem}[section]
\newtheorem{corollary}{Corollary}[theorem]
\newtheorem{remark}{Remark}
\newtheorem{lemma}[theorem]{Lemma}
\usepackage{amsmath}    
  {
      \theoremstyle{plain}
      \newtheorem{assumption}{Assumption}
  }

\theoremstyle{definition}

\usepackage{xparse}%
\usepackage{blindtext}
\usepackage{morewrites}
\usepackage{wrapfig}
\usepackage{xkeyval}%

\newwrite\authorbibfile%

\AtBeginDocument{%
  \immediate\openout\authorbibfile=\jobname.aub%
}%

\AtEndDocument{%
\immediate\closeout\authorbibfile
\InputIfFileExists{\jobname.aub}{}{}
}%

\makeatletter

\define@key{authorbib}{scale}[1]{%
\def\AuthorbibKVMacroScale{#1}%
}

\define@key{authorbib}{wraplines}[10]{%
\def\AuthorbibKVMacroWraplines{#1}%
}

\define@key{authorbib}{imagewidth}[4cm]{%
\def\AuthorbibKVMacroImagewidth{#1}%
}

\define@key{authorbib}{overhang}[10pt]{%
\def\AuthorbibKVMacroOverhang{#1}%
}

\define@key{authorbib}{imagepos}[l]{%
\def\AuthorbibKVMacroImagepos{#1}%
}

\makeatother

\presetkeys{authorbib}{imagepos=l,imagewidth=4cm,wraplines=15,overhang=20pt}{}

\newlength{\AuthorbibTopSkip}
\newlength{\AuthorbibBottomSkip}
\NewDocumentCommand{\authorbibliography}{+o+m+m+m}{%
  \IfNoValueTF{#1}{%
  }{%
    \setkeys{authorbib}{#1}%
    \immediate\write\authorbibfile{%
      \string\begin{wrapfigure}[\AuthorbibKVMacroWraplines]{\AuthorbibKVMacroImagepos}[\AuthorbibKVMacroOverhang]{\AuthorbibKVMacroImagewidth}^^J
        \string\includegraphics[scale=\AuthorbibKVMacroScale]{#2}^^J
        \string\end{wrapfigure}^^J
    }%
  }%
  \IfNoValueTF{#3}{%
    \typeout{Warning: No author name}%
  }{%
    \immediate\write\authorbibfile{%
      \unexpanded{\vspace{\AuthorbibTopSkip}}^^J
      \string\noindent\relax
      \unexpanded{\textbf{#3}\par}^^J
      \string\noindent\relax
      \unexpanded{#4}^^J%
      \unexpanded{\vspace{\AuthorbibBottomSkip}}^^J
      }%
  }%
}%

\begin{document}

\begin{frontmatter}

\title{Distributed Localization in Dynamic Networks via Complex Laplacian} 

\thanks[footnoteinfo]{
This work was supported 
by Nanyang Technological University under the Wallenberg-NTU Presidential Postdoctoral Fellowship and Projects of Major International (Regional) Joint Research Program under NSFC Grant no. 61720106011 and 62103352. (Corresponding author: Lihua Xie.)
}

\author[ntu]{Xu Fang}\ead{fa0001xu@e.ntu.edu.sg},    
\author[ntu]{Lihua Xie\thanksref{footnoteinfo}}\ead{elhxie@ntu.edu.sg},
\author[ntu]{Xiaolei Li}\ead{xiaolei@ysu.edu.cn}

\address[ntu]{School of Electrical and Electronic Engineering, Nanyang Technological University, 639798,  Singapore}                                             


\begin{abstract}            
Different from most existing distributed localization approaches in static networks where the agents in a network are static,
this paper addresses the distributed localization problem in dynamic networks where the positions of the agents are time-varying. Firstly, 
complex constraints for the positions of the agents are constructed based on local relative position (distance and local bearing) measurements.
Secondly, both algebraic condition and graph condition of network localizability in dynamic networks are given. Thirdly, a distributed localization protocol is proposed such that all the agents can cooperatively find their positions by solving the complex constraints in dynamic networks.
Fourthly, the proposed method is extended to address the problem of integrated distributed localization and formation control. It is worth mentioning that the proposed algorithm can also be applied in the
case that only distance and sign of direction measurements are available, where the sign of direction measurement is a kind of one bit local relative measurement and has less information than local bearing. 
\end{abstract}
\end{frontmatter}

\section{Introduction}

Scientist around the world are recognizing the benefits of autonomous agents (unmanned ground vehicles and unmanned aerial vehicles) in both civilian and military applications such as cooperative transportation, geographical
data collection, and cooperative search and rescue \citep{letizia2021novel,lu2021resource,wu2020cooperative}. Enabling such applications requires a reliable localization system for the agents to achieve self-localization. 

Network localization aims to estimate the positions of all agents given the positions of part of the agents and inter-agent relative measurements \citep{laoudias2018survey,aspnes2006theory,buehrer2018collaborative}. The centralized network localization
requires a central unit to process the information of all agents, where the positions of all agents are calculated in a centralized way \citep{fang2020graph,schmuck2019ccm,wang2017ultra}. But 
in a large network, it is impractical to use a central unit to calculate the positions of all agents \citep{fang2020m,priyantha2000cricket}. Hence, each agent in a large network should have the ability to achieve self-localization in a distributed manner by performing local computations based on available relative measurements with its neighbors.

According to different types of inter-agent relative measurements, most of the existing approaches to distributed network localization can be divided into three categories:  (1) distance measurement based \citep{eren2004rigidity,diao2014barycentric,khan2009distributed}; (2) bearing measurement based \citep{shames2012analysis,cao2021bearing,zhao2016localizability}; (3) angle measurement based \citep{jing2019angle1,lin2017mix}. The existing distributed localization approaches \citep{eren2004rigidity,shames2012analysis,cao2021bearing,zhao2016localizability,diao2014barycentric,khan2009distributed,jing2019angle1,lin2017mix} mainly focus on static networks, i.e., the agents in a network are static. The problem becomes challenging when it comes to distributed localization in dynamic networks, where the positions of the agents are time-varying such that the network localizability is difficult to be guaranteed. For instance, each agent in static networks requires three non-colinear neighbors to achieve self-localization in 2-D distance-based distributed localization \citep{eren2004rigidity,diao2014barycentric,khan2009distributed}.
But in dynamic networks, it is difficult to guarantee the positions of its 
three neighbors to be non-colinear at any time instant. Similar issues can also be found in 2-D bearing-based or angle-based distributed localization \citep{shames2012analysis,cao2021bearing,zhao2016localizability,jing2019angle1,lin2017mix}, where each agent and its two neighbors must be non-colinear or requires three non-colinear neighbors.

The distributed localization in dynamic networks is worth studying in multi-agent networks as the agents need to change their positions to keep specific formation or cover a large area. 
Different from the above existing distributed localization methods in static networks, we aim to propose a distributed localization protocol in 2-D dynamic networks based on local relative positions (distance and local bearing), where each agent can measure the relative positions in its local coordinate frame with unknown orientation. In addition, we aim to apply the distributed localization technique to formation control, i.e, we also study integrated distributed localization and formation control.

Two fundamental problems, namely network localizability and distributed protocols in dynamic networks, will be studied in this article. Similar to static networks, we also need to explore the position relationship among the agents in dynamic networks based on the relative measurements. In this paper, we use
complex 
constraints to describe the position relationship among the agents. A remarkable advantage of the complex 
constraints is that
the non-colinear condition is removed, e.g., each agent does not require three non-colinear neighbors or each agent can be colinear with its two neighbors. Different from existing relative-position-based 
distributed localization approaches \citep{fang2020,barooah2007estimation,lin2015distributed,stacey2017role,mendrzik2020joint,oh2013formation} which require the global coordinate frame, orientation alignment, or can only be applied in static networks, this paper studies distributed localization in dynamic networks based on local relative positions, where the local coordinate frames of different agents can have different unknown time-varying orientations.
Moreover, compared with local-relative-position-based distributed localization \cite{fang2020} which requires each agent to have three neighbors, each agent in this article only needs two neighbors.

Although there are some results on complex-Laplacian-based distributed localization \citep{lin2015distributed,lin2016distributedb} and formation control \citep{lin2014distributed,han2015formation,de2021distributed}, they need undirected graphs \citep{lin2016distributedb,de2021distributed}, aligned coordinate frame and synchronized velocities of the leaders \citep{han2015formation}, or can only be applied in static networks \citep{lin2015distributed,lin2016distributedb}. These constraints are removed in our proposed complex-Laplacian-based distributed localization and formation.
In addition, different from existing complex-Laplacian-based approaches which require relative position (distance and bearing) measurements \citep{lin2014distributed,lin2015distributed,han2015formation,de2021distributed}, we will also explore the more challenging case that only distance and sign of direction measurements are available, where the sign of direction measurement is a kind of one bit local relative measurement and has less information than local bearing. The main contributions of this article are summarized as following:
\begin{enumerate}[(1)]
    \item Complex constraints for the positions of the agents in dynamic networks are constructed based on local relative positions, where each agent only needs two neighbors. 
    \item Both algebraic condition and graph condition are given to examine whether a dynamic network is localizable. 
    \item A distributed localization algorithm is proposed such that all agents can cooperatively find their positions by solving the complex constraints in dynamic networks.
    \item The proposed method is extended to integrated distributed localization and formation control, and is also applied to the case when only distance and sign of direction measurements are available. 
\end{enumerate}

The rest of the paper is organized as follows. The notations and problem statement are given in Section \ref{problem}. Section \ref{displace} introduces the complex constraint and its invariance property. The algebraic condition and graph condition are explored in Section \ref{secdyna} to guarantee the localizability of dynamic networks. Section \ref{protocol} proposes a distributed localization protocol in dynamic networks to estimate the unknown positions of the agents based on local relative positions, which is then extended to the case when only distance and sign of direction measurements are available in Section \ref{exte}. Section \ref{formation} shows that the proposed method can also be extended to integrated distributed localization and formation control. A simulation example is given in Section \ref{simul} and the conclusion is drawn in Section \ref{conc}.

\begin{figure*}[t]
\centering
\includegraphics[width=0.8\linewidth]{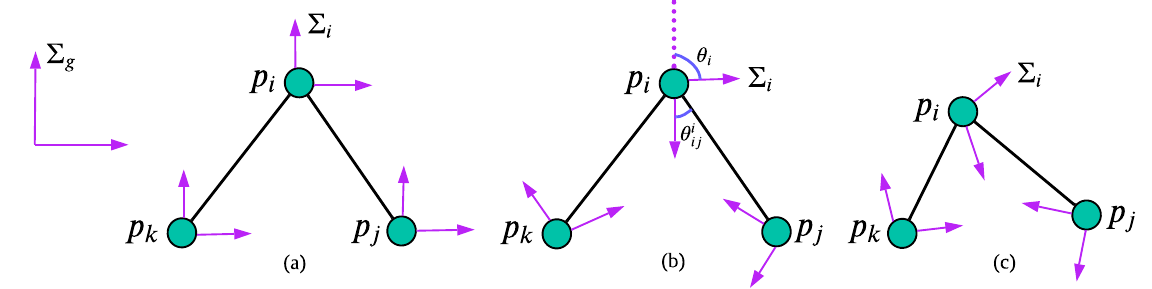}
\caption{(a) Relative positions in global coordinate frame. (b) A configuration with relative positions in local coordinate frames. (c) Different local coordinate frames lead to a different configuration with same local relative position constraints as in (b).}
\label{sign}
\end{figure*}

\section{Notations and Problem Statement}\label{problem}

\subsection{Notations}

Let $\mathbb{R}$ and $\mathbb{C}$ be the set of real and complex numbers, respectively.  ${I}_d \in \mathbb{R}^{d \times d}$ represents an identity matrix of dimension $d \times d$, while ${\mathbf{0}}_d \in \mathbb{R}^{d}$ represents a vector with all entries equal to zero.
Let $\text{det}(\cdot)$ and $\text{rank}(\cdot)$ be the determinant and rank of a matrix, respectively. Consider a communication graph $\mathcal{G}=\{ \mathcal{V},\mathcal{E}\}$, where $\mathcal{V}$ and $\mathcal{E} \subseteq \mathcal{V} \times \mathcal{V}$ are the agent set and edge set, respectively. If $(i,j) \in \mathcal{E}$, agent $j$ is a neighbor of agent $i$.  $\mathcal{N}_i \!=\! \{ j \! \in \! \mathcal{V}  |  (i,j) \in \mathcal{E} \}$ represents the neighbor set of agent $i$. Denote $\mathcal{U}: v_1 \rightarrow v_2 \rightarrow \cdots \rightarrow v_k$ as a walk from vertex $v_1$ to vertex $v_k$. The walk $\mathcal{U}$ is called a path if the vertexes are distinct. A path is called a cycle if 
the starting vertex and ending vertex are the same and all other vertexes are distinct. 
A vertex $v \in \mathcal{V}$ is called two-reachable from a set $P$ if there are two disjoint paths from $P$ to $v$. Let $\mathcal{A}^T$ be the transpose of a real matrix $\mathcal{A} \in \mathbb{R}^{d \times d}$. Let $\mathcal{A}^H$ be the conjugate transpose of a complex matrix $\mathcal{A} \in \mathbb{C}^{d \times d}$. Let $\Sigma_g$ be a common global coordinate frame of the complex plane.
The position $p_i$ of each agent $i$ in $\Sigma_g$ shown in Fig. \ref{sign}(a) is denoted by 
\begin{equation}\label{complex}
    p_i = x_i + y_i \iota,
\end{equation}
where $x_i, y_i \in \mathbb{R}$ and
$\iota$ is the imaginary unit with $\iota^2 \!=\! -1$.  $p_i^H= x_i - y_i \iota$ is the complex conjugate of $p_i$.  Let $\| \cdot \|_2$ be the $\mathcal{L}_2$ norm and $\| p_i \|_2 = \sqrt{x_i^2+y_i^2}$. Let $\Sigma_i$ be the 
local coordinate frame of agent $i$ shown in Fig. \ref{sign}(a)-(c).
Let $\theta_i \in [0, 2\pi]$ be the unknown orientation difference between $\Sigma_i$ and $\Sigma_g$ shown in Fig. \ref{sign}(b).  Define
\begin{equation}\label{ek1}
e_{ij} = p_j-p_i, \ e_{ij}^{i} = e_{ij} \cdot \text{exp}(\theta_i \iota),
\end{equation}
where  $\text{exp}(\cdot)$ is the natural exponential function. $e_{ij}$ and $e_{ij}^{i}$ are the relative position in global coordinate frame $\Sigma_g$ and local coordinate frame $\Sigma_i$, respectively. The local relative position $e_{ij}^{i}$ can be obtained by the distance $d_{ij}$ and local relative bearing $\theta^i_{ij} \in [0, 2\pi]$ shown in Fig. \ref{sign}(b). 
\begin{equation}
 e_{ij}^{i} = d_{ij} \cdot \text{exp}(\theta^i_{ij} \iota), \ d_{ij}=\| p_j-p_i \|_2. 
\end{equation}

The cross product $\times$ of the complex numbers $p_i$ and $p_j$ is defined as
\begin{equation}\label{cross}
    p_i \times p_j = x_iy_j -x_jy_i.
\end{equation}

Let $(\mathcal{G}, p)$ be a network of $n$ agents, where $p=(p_1^H, \cdots, p_n^H)^H$ is the configuration of the network. 

\begin{defn}
A network $(\mathcal{G}, p(t))$ is called a dynamic network if the positions of the agents $p(t)$ are time-varying. A dynamic network $(\mathcal{G}, p(t))$ is also called a formation if it is applied to multi-agent systems. 
\end{defn}

\subsection{Problem Statement}

In this paper, we consider the distributed localization in dynamic network $(\mathcal{G}, p)$ with a group of $m$ leaders and $n \!-\! m$ followers in 2-D plane.  The leader set and follower set are represented by $\mathcal{V}_l=\{1, \cdots, m \}$ and $\mathcal{V}_f\!=\! \{m\!+\!1, \cdots, n\}$, respectively.  The positions of the leaders and followers are given by $p_l \!=\! [p_1^H, \cdots, p_{m}^H]^H$ and $p_f \!=\! [p^H_{m\!+\!1} , \cdots, p^H_{n}]^H$, respectively. 
Denote $\hat p_i$ as a position estimate of agent $i$. We aim to design a position estimator for each follower given the  positions of the leaders and inter-agent local relative positions (distance and local bearing) or inter-agent local distance and sign of direction measurements. That is, 
\begin{equation}\label{con1l}
\begin{array}{ll}
      \lim\limits_{t \rightarrow \infty} (\hat p_i(t)-p_i(t)) = 0, \ \ i \in \mathcal{V}_f.
\end{array}
\end{equation}

\begin{remark}
As stated in  \cite{lin2015distributed}, the local-relative-position-based distributed localization problem is not trivial. Since the orientation difference $\theta_i$ in \eqref{ek1} is unknown, we cannot calculate global relative position $e_{ij}$ based on local relative position $e_{ij}^{i}$.

\end{remark}

Next, we will introduce the tools to achieve distributed localization in dynamic networks.

\section{Complex Constraint}\label{displace}

\subsection{Complex Constraint}\label{displacon}

Agent $i$ and agent $j$ are called collocated if $p_i = p_j$.
When agent $i$ and its two neighboring agents $j,k$ in $\mathbb{C}$ are not collocated, we can obtain a complex constraint, i.e.,
\begin{equation}\label{complexc}
    w_{ij}e_{ij}+w_{ik}e_{ik}= 0, 
\end{equation}
where $w_{ij}, w_{ik} \in  \mathbb{C}$ are the complex weights designed as
\begin{equation}\label{weight}
 w_{ij} = \frac{e_{ij}^H}{d_{ij}^2}, \ w_{ik} = -\frac{e_{ik}^H}{d_{ik}^2}. 
\end{equation}

Equation \eqref{complexc} can be rewritten as
\begin{equation}\label{complexc1}
    w_{ii}p_i = w_{ij} p_j + w_{ik}p_k,
\end{equation}
where $w_{ii}=w_{ij}+w_{ik}$. 

\begin{lemma}\label{le1}
(\textbf{Property of Complex Constraint}) If there is no inter-collision among each agent $i$ and its two neighboring agents $p_j,p_k$, the complex constraint \eqref{complexc} with its complex weights \eqref{weight} has the following properties: 
\begin{enumerate}[(1)]
    \item The parameters $w_{ij}, w_{ik} \neq 0$;
    \item The parameter $w_{ii} \neq 0$. 
\end{enumerate}
\end{lemma}

\textbf{Proof.}
(1) Since $e_{ij}, e_{ik} \neq 0$, it yields from \eqref{weight} that $w_{ij}=\frac{e_{ij}^H}{d_{ij}^2} \neq 0$ and $w_{ik} = -\frac{e_{ik}^H}{d_{ik}^2} \neq 0$. (2)  If $w_{ii}=0$, we have $w_{ij}=-w_{ik}$. Then,
it yields from \eqref{complexc1} that $w_{ij}(p_j-p_k)=0$, i.e., $p_j=p_k$, which contradicts the fact that $p_j \neq p_k$. Thus, $w_{ii} \neq 0$. \ \ \ \ \ \ \ \ \ \ \ \ \ \ \ \ \ \ \ \  \ \ \ \ \ \ \ \ \ \ \ \ \ \ \ \ \ \ \ \ \ \ \ \ \ \ \ \ 
 \qed

It is clear from Lemma \ref{le1} that
$p_i$ can be localized by its two neighbors $p_j$ and $p_k$ if there is no inter-collision among each agent $i$ and its two neighboring agents $p_j,p_k$, i.e., 
\begin{equation}\label{complexc2}
 p_i = \frac{w_{ij}}{w_{ii}} p_j +  \frac{w_{ik}}{w_{ii}}p_k.
\end{equation}

\subsection{Invariance of Complex Constraint}\label{inva}

Let $p'_i, p'_j, p'_k$ be
\begin{equation}
\begin{array}{ll}
     &     p'_i \!=\! \beta +  p_i \cdot \text{exp}(\theta\iota), \\
     & p'_j \!=\! \beta + p_j \cdot \text{exp}(\theta\iota), \\
     & p'_k \!=\! \beta + p_k \cdot \text{exp}(\theta\iota),
\end{array}
\end{equation}
where $\beta \in \mathbb{C},  \theta \in [0, 2\pi]$ are translation and rotation parameters, respectively. We have
\begin{equation}\label{eo1}
\begin{array}{ll}
     &  e'_{ij}\!=\!p'_j-p'_i =  e_{ij} \cdot \text{exp}(\theta\iota),  \\
     & e'_{ik}\!=\!p'_k-p'_i = e_{ik} \cdot \text{exp}(\theta\iota).
\end{array}
\end{equation}

Then, we obtain
\begin{equation}\label{invg}
    w_{ij}e'_{ij} + w_{ik} e'_{ik} =  (w_{ij}e_{ij} + w_{ik} e_{ik}) \cdot \text{exp}(\theta\iota).
\end{equation}

It yields from \eqref{invg} that the complex constraint \eqref{complexc} is invariant to translations and rotations of $p_i ,p_j, p_k$, i.e.,
\begin{equation}\label{insp}
w_{ij}e_{ij} + w_{ik} e_{ik} = 0 \Longleftrightarrow w_{ij}e'_{ij} + w_{ik} e'_{ik} = 0.    
\end{equation}

\subsection{Local Relative Position based Complex Constraint}\label{localco}

Denote 
$e_{ij}^i, e_{ik}^i$ as
the local relative positions measured by agent $i$ in $\Sigma_i$. From \eqref{ek1}, we have
\begin{equation}
 e_{ij}^{i} = e_{ij} \cdot \text{exp}(\theta_i \iota), \ e_{ik}^{i} = e_{ik} \cdot \text{exp}(\theta_i \iota).
\end{equation}

Although the orientation difference $\theta_i$ between $\Sigma_i$ and $\Sigma_g$ is unknown, we can know from \eqref{insp} that
\begin{equation}
w_{ij}e_{ij} + w_{ik} e_{ik} = 0 \Longleftrightarrow w_{ij}e_{ij}^{i}  + w_{ik} e_{ik}^{i} = 0.       
\end{equation}

Hence, the complex weights $w_{ij}, w_{ik}$ in \eqref{complexc} can be calculated based on local relative positions $e_{ij}^i, e_{ik}^i$ by \eqref{weight}, i.e.,
\begin{equation}\label{weightl}
w_{ij} = \frac{(e_{ij}^i)^H}{\|e_{ij}^i\|_2^2}, \ w_{ik} = -\frac{(e_{ik}^i)^H}{\|e_{ik}^i\|_2^2}. 
\end{equation}

\section{Localizability of Dynamic Network}\label{secdyna}

As shown in Section \ref{localco},
the complex constraint $w_{ij}e_{ij} \!+\! w_{ik}e_{ik} \!= \!0,  j, k \! \in \! \mathcal{N}_i$ can be obtained by local relative positions.

\begin{defn}\label{defn1}
The dynamic network $(\mathcal{G}, p(t))$ is called lcoalizable if the positions of the followers $p_f(t)$ can be uniquely expressed by the positions of the leaders $p_l(t)$ given the inter-agent complex constraints $w_{ij}e_{ij}+w_{ik}e_{ik}=0, j, k \in \mathcal{N}_i$. 
\end{defn}

\begin{assumption}\label{asu1}
Each follower has at least two neighbors. 
\end{assumption}

\begin{assumption}\label{asu11}
No two agents are collocated.
\end{assumption}

\begin{remark}
Assumption \ref{asu11} can be removed by designing appropriate tracking controllers of the agents shown in Section \ref{formation}.
\end{remark}

\subsection{Algebraic Condition for Localizability}

Under Assumption \ref{asu1}, each follower $i \in \mathcal{V}_f$ can form a complex constraint \eqref{complexc} based on local relative positions. Then, we have
\begin{equation}\label{expre}
 w_{ii}p_i = w_{ij} p_j + w_{ik}p_k,
\end{equation}
where $w_{ii} = w_{ij} \!+\! w_{ik}$. We can aggregate the linear complex constraint of each follower from \eqref{expre} into a matrix form, i.e.,
\begin{equation}\label{g1}
    W_f p = \mathbf{0},
\end{equation}
where $p=(p_1^H, \cdots, p_n^H)^H$ and
\begin{equation}\label{zer1}
\begin{array}{ll}
    &[W_f]_{ij} \!=\!  \left\{ \! \begin{array}{lll} 
    -w_{ij}, & i \in \mathcal{V}_f, \
     j \in \mathcal{N}_i, \ j \neq i, \\
    \ \  0, &   i \in \mathcal{V}_f, \ j \notin \mathcal{N}_i, \ j \neq i, \\
    \sum\limits_{k \in \mathcal{N}_i} w_{ik},
    & i \in \mathcal{V}_f, \ j=i. \\
    \end{array}\right. 
\end{array} 
\end{equation}

Note that \eqref{g1} can be rewritten as
\begin{equation}\label{loi}
   W_{fl} p_l +    W_{f\!f} p_f = \mathbf{0},
\end{equation}
where $p_l \!=\! [p_1^H, \cdots, p_{m}^H]^H$ and $p_f \!=\! [p^H_{m\!+\!1} , \cdots, p^H_{n}]^T$ are the positions of the leaders and followers, respectively. $ W_f \!=\![
W_{fl} \  W_{f\!f} 
]$ is a complex matrix with $ W_{fl} \in \mathbb{C}^{(n\!-\!m)\times m}$ and $W_{f\!f} \in \mathbb{C}^{(n\!-\!m)\times (n\!-\!m)}$. 
Next, we will explore that under what conditions, the dynamic network $(\mathcal{G}, p(t))$ is localizable. 

\begin{theorem}\label{alge}
(\textbf{Algebraic Condition}) Under Assumptions \ref{asu1}-\ref{asu11},
a dynamic network $(\mathcal{G}, p(t))$ is localizable if and only if the complex matrix $W_{f\!f}(t)$ in \eqref{loi} is invertible.
\end{theorem}

\textbf{Proof.} It is straightforward from Definition \ref{defn1} and \eqref{loi} that  a dynamic network $(\mathcal{G}, p(t))$ is localizable if and only if the complex matrix $W_{f\!f}(t)$ is invertible. Then, the positions of the followers $p_f(t)$ can be uniquely expressed by the positions of the leaders $p_l(t)$, i.e.,
\begin{equation}
 p_f(t) = -W_{f\!f}^{-1}(t)W_{fl}(t)p_l(t).
\end{equation}
\ \ \ \ \ \ \ \ \ \ \ \ \ \ \ \ \ \ \ \  \ \ \ \ \ \ \ \ \ \ \ \ \ \ \ \ \ \ \ \ \ \ \ \ \ \ \ \ \ \ \ \ \ \ \ \ \ \ \ \ \ \ \ \ \ \ \ \ \ \ \ \ \ \ 
 \qed

\begin{figure}[t]
\centering
\includegraphics[width=1\linewidth]{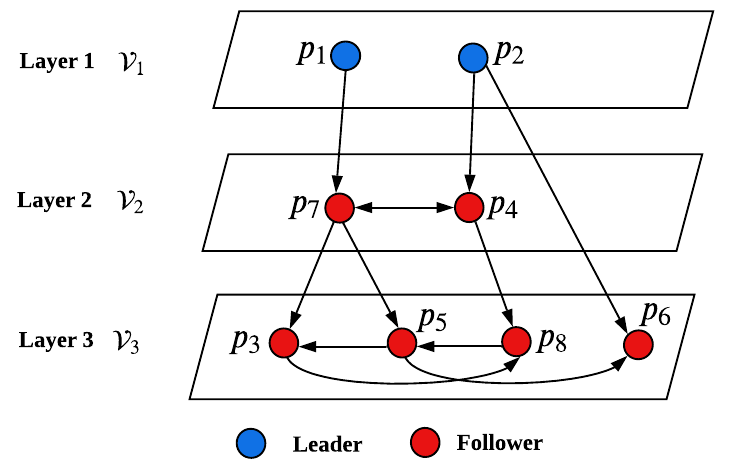}
\caption{A $\kappa$-layer graph.}
\label{nece5}
\end{figure}

\subsection{Graph Condition for Localizability}

Before exploring the graph condition for localizability, we first introduce the concept of $\kappa$-layer graph.

\begin{defn}\label{fn3}
A graph $\mathcal{G}$ is called a $\kappa$-layer graph if
\begin{enumerate}[(1)]
\item The agents in $\mathcal{G}$ are divided into $\kappa \!>\! 1$ subsets $\mathcal{V}_1,  \mathcal{V}_{2},$ $\cdots, \mathcal{V}_{{\kappa}}$, where $\mathcal{V}_i \cap \mathcal{V}_j = \emptyset$ if $i \neq j$. Agent $i$ is called in layer $h_i$ if  $i \in \mathcal{V}_{h_i} (1 \le h_i \le \kappa)$;
\item  Subset $\mathcal{V}_1$ includes all leaders, i.e., $\mathcal{V}_1=\mathcal{V}_l$. The union of the subsets $\mathcal{V}_{2}, \cdots, \mathcal{V}_{{\kappa}}$ include all followers, i.e.,  $\bigcup\limits_{s=2}^{\kappa} \mathcal{V}_s = \mathcal{V}_f$;
\item If follower $i$ of layer $h_i$ is in a cycle $\mathcal{C}$, all agents in cycle $\mathcal{C}$ are two-reachable from another two agents $k,g$,
where agents $k,g$ are in the layers $h_k, h_g < h_i$ and
each agent in cycle $\mathcal{C}$ has access to either agent ${k}$ or agent ${g}$. The agents in a cycle $\mathcal{C}$ are not required to be in the same layer;
\item If follower $i$ of layer ${h}_i$ does not belong to any cycle, its neighbor $j \in \mathcal{N}_i$ is either  in a lower layer $h_j(h_j<h_i)$ or in a cycle with $h_j(h_j \le h_i)$. 
\end{enumerate}
\end{defn}

\begin{remark}
The $\kappa$-layer graph is
a kind of hierarchical-decomposition-based graph \citep{wang2013hierarchical,wang2015hierarchical}, where conditions (1)-(4) in Definition \ref{fn3} can be regarded as a way of hierarchical decomposition. In our case, the leaders are in layer $1$ and the followers
are in the rest layers.
The number of followers in each layer
can be randomly chosen, and each follower can be assigned to any layer. It is proved in Theorem \ref{trem1} that each follower $i$ can be localized by its neighbors in dynamic networks if either condition (3) or condition (4) holds.
\end{remark}

\begin{remark}
 The $\kappa$-layer graph $\mathcal{G}$ allows both directed and undirected edges, and also allows the existence of the cycles. Thus, 
the $\kappa$-layer graphs are more relaxed than the widely used directed acyclic graphs in distributed localization and formation control of multi-agent systems \citep{cao2019relative,ding2010collective}. 
\end{remark}

A simple example of a $\kappa$-layer graph is given in Fig. \ref{nece5}, whose details are given below.
\begin{enumerate}[(1)]
\item The agents are divided into three layers $1,  {2}, {{3}}$, where the leaders $p_1,p_2$ are in layer $1$ and the followers $p_3, \cdots, p_8$ are in layer $2$ and layer $3$;

\item  Follower $p_{3}$ of layer $3$ is in
a circle $\mathcal{C}: p_{3} \rightarrow p_{8} \rightarrow p_{5} \rightarrow p_{3}$. All agents in circle $\mathcal{C}$ are two reachable from another two agents $p_4, p_7$ of layer $2$, where each agent in circle $\mathcal{C}$ has access to either $p_4$ or $p_7$.

\item Follower $p_{6}$ of layer $3$ does not belong to any circle and has two neighbors $p_{2},p_{5}$, where
$p_{2}$ is in a lower layer $1$ and $p_{5}$ is in a cycle of layer $3$. 

\end{enumerate}

\begin{figure}[t]
\centering
\includegraphics[width=1\linewidth]{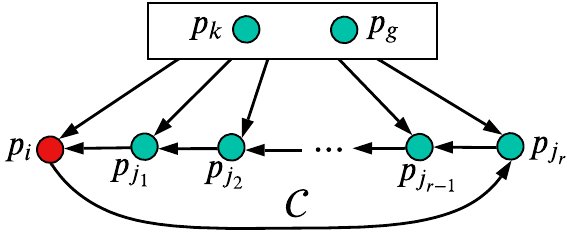}
\caption{Follower $i$ of layer $h_i$ is in a cycle $\mathcal{C}$. Agent $k$ is in layer $h_k$ and agent $g$ is in layer $h_g$, where $h_k,h_g<h_i$. }
\label{nece7}
\end{figure}

\begin{theorem}\label{trem1}
(\textbf{Graph Condition for Localizability})  Under Assumptions \ref{asu1}-\ref{asu11}, 
a dynamic network  $(\mathcal{G}, p(t))$ with a $\kappa$-layer graph is localizable.
\end{theorem}

\textbf{Proof.} 
In a $\kappa$-layer graph, 
we will first  prove that the position of each follower $i$ in layer $h_i(2 \le h_i \le \kappa)$ can be expressed by the positions of the agents in the lower layers through elementary row operations of the matrix $W_{f}$ \eqref{g1}.

\textbf{Case (\romannumeral1)}. If follower $i$ of layer $h_i$ is in a cycle $\mathcal{C}$ shown in Fig. \ref{nece7}, there
are $r \!+\!1$ agents $i, j_1, j_2, \cdots, j_r$ in the cycle $\mathcal{C}$.
We can know from Definition \ref{fn3} that all
agents $i, j_1, j_2, \cdots, j_r$ in the cycle $\mathcal{C}$ are two-reachable from another two agents $k,g$, where agents $k,g$ are in the layers lower than layer $h_i$ and each agent in cycle $\mathcal{C}$ has access to either agent $k$ or agent $g$. Note that each agent can form a complex constraint \eqref{complexc} with its two neighbors. Thus, for the agents $i,j_1,j_2,\cdots,j_r$ in cycle $\mathcal{C}$, 
it yields from \eqref{expre} that
\begin{equation}\label{h1}
\begin{array}{ll}
    &  \left\{ \! \begin{array}{lll} 
       w_{ii}p_i = w_{ij_1}p_{j_1} + w_{ib_0}p_{b_0},  \\
        w_{j_1j_1}p_{j_1} = w_{j_1j_2}p_{j_2} + w_{j_1b_1}p_{b_1}, \\
           \cdots \\
       w_{j_{r\!-\!1}j_{r\!-\!1}}p_{j_{r\!-\!1}} = w_{j_{r\!-\!1}j_{r}}p_{j_r} + w_{j_{r\!-\!1}b_{{r\!-\!1}}}p_{b_{r\!-\!1}},
    \end{array}\right.
\end{array} 
\end{equation}
and 
\begin{equation}\label{h2}
w_{j_rj_r}p_{j_r} = w_{j_ri}p_{i} + w_{j_rb_{{r}}}p_{b_{r}},    
\end{equation}
where $b_0,b_1,\cdots,b_r \in \{ k,g\}$. Under Assumption \ref{asu11}, no two agents are collocated. Then, we can know from Lemma \ref{le1} that the parameters $w_{ii},w_{j_1j_1}, \cdots,w_{j_rj_r} \neq 0$. From \eqref{h1}, we obtain
\begin{equation}\label{h3}
    w_{ii}p_i =  w_{ib_0}p_{b_0}+  \sum\limits_{l=1}^{r-1} w_{ib_l}p_{b_l} + w_{ij_r}p_{j_r},
\end{equation}
where $w_{ii} \!=\! w_{ib_0}\!+\! \sum\limits_{l=1}^{r\!-\!1} w_{ib_l}\!+\!w_{ij_r}$, $w_{ij_r} \!=\! w_{ij_1} \prod\limits_{s=1}^{r\!-\!1} \frac{w_{j_sj_{s\!+\!1}}}{w_{j_sj_{s}}}$, and
\begin{equation}
\begin{array}{ll}
    &w_{ib_l} \!=\!  \left\{ \! \begin{array}{lll} 
    \frac{w_{ij_1}w_{j_1b_1}}{w_{j_1j_1}}, & l=1, \\
  \frac{w_{ij_1}w_{j_lb_l}}{w_{j_lj_l}} \prod\limits_{s=1}^{l-1} \frac{w_{j_sj_{s\!+\!1}}}{w_{j_sj_{s}}}, &  l>1. \\
    \end{array}\right. 
\end{array} 
\end{equation}

Since $b_0,b_1,\cdots,b_{r\!-\!1} \in \{ k,g\}$, \eqref{h3} can be rewritten as
\begin{equation}\label{h4}
 w_{ii}p_i =  w_{ik}p_k + w_{ig}p_g + w_{ij_r}p_{j_r},
\end{equation}
where $w_{ii}=w_{ik}+w_{ig}+w_{ij_r}$. Under Assumption \ref{asu11} and from Lemma \ref{le1}, we have $w_{ii} \neq 0$ and $p_i \neq p_{j_r}$. Then, it is concluded from \eqref{h4} that $\|w_{ik} \|_2 + \|w_{ig}\|_2 \neq 0$. Combining \eqref{h2} and \eqref{h4}, we obtain
\begin{equation}\label{h5}
w_{ii}p_i = w_{ik}p_k + w_{ig}p_g + w_{ij_r}\frac{ w_{j_ri}}{ w_{j_rj_r}}p_i + w_{ij_r}\frac{ w_{j_rb_r}}{ w_{j_rj_r}}p_{b_r}. 
\end{equation}

Since $b_r \in \{ k,g\}$, 
\eqref{h5} can also be rewritten as
\begin{equation}\label{gh3}
\bar w_{ii} p_i =  \bar w_{ik}p_k + \bar w_{ig}p_g,
\end{equation}
where  $\bar w_{ii} = \bar w_{ik} + \bar w_{ig}$ and
\begin{equation}\label{fc}
    \begin{array}{ll}
    &  \left\{ \! \begin{array}{lll} 
      \bar w_{ii} = w_{ii} - w_{ij_r}\frac{ w_{j_ri}}{ w_{j_rj_r}}, \\
        \bar w_{ik} = w_{ik}+ w_{ij_r}\frac{ w_{j_rb_r}}{ w_{j_rj_r}}, \bar w_{ig} = w_{ig}  \ \text{if} \ b_r=k, \\
           \bar w_{ik} = w_{ik}, \bar w_{ig} = w_{ig} + w_{ij_r}\frac{ w_{j_rb_r}}{ w_{j_rj_r}}  \ \text{if} \ b_r=g.
    \end{array}\right.
\end{array} 
\end{equation}

Since $p_k \neq p_g$ and $\| w_{ik}\|_2 \!+\! \|w_{ig} \|_2\neq 0$, it is concluded from \eqref{gh3} and \eqref{fc} that $\bar w_{ii} \neq  0$.
Hence, 
follower $i$ can be expressed by the agents $k,g$, i.e.,
\begin{equation}\label{h6}
    p_i = \frac{\bar w_{ik}}{\bar w_{ii}}p_k + \frac{\bar w_{ig}}{\bar w_{ii}}p_g. 
\end{equation}

The calculations in \eqref{h3}-\eqref{h6} can be achieved through elementary row operations of the matrix $W_{f}$ \eqref{g1}.
Since the agents $k,g$ are in the layers lower than layer $h_i$, we can know from \eqref{h6} that
if follower $i$ of layer $h_i$ is in a cycle, its position can be expressed by the positions of the agents in the lower layers through elementary row operations of the matrix $W_{f}$ \eqref{g1}.

\textbf{Case (\romannumeral2)}. If follower $i$ of layer $h_i$ is not in a cycle,  it yields from \eqref{expre} that
\begin{equation}
 w_{ii}p_i = w_{ij} p_j + w_{ik}p_k,   \ j,k \in \mathcal{N}_i. 
\end{equation}

Under Assumption \ref{asu11}, we can know from Lemma \ref{le1} that the parameters $w_{ii} \neq 0$. Then, we have
\begin{equation}\label{h7}
    p_i = \frac{w_{ij}}{w_{ii}}p_j + \frac{w_{ik}}{w_{ii}}p_k.
\end{equation}

From Definition \ref{fn3}, we can know 
that the agents $j,k \in \mathcal{N}_i$
are either in the lower layers $h_j, h_k <h_i$ or in the cycles with $h_j , h_k \le h_i$. From the above Case (\romannumeral1) and \eqref{h7}, we can know that if follower $i$ of layer $h_i$ is not in a cycle, its position can also be expressed by the positions of the agents in the lower layers through elementary row operations of the matrix $W_{f}$ \eqref{g1}.

From the above  Case (\romannumeral1) and  Case (\romannumeral2), the position of each follower $i$ in layer $h_i$ can be expressed by the positions of the agents in the lower layers through elementary row operations of the matrix $W_{f}$ \eqref{g1}. Thus, the position of each follower $i$ in layer $h_i$ can be finally expressed by the positions of the leaders in layer $1$ through elementary row operations of the matrix $W_{f}$, i.e.,
\begin{equation}\label{fin}
  p_i = \sum\limits_{j \in \mathcal{V}_1} \alpha_{ij}p_j, \ i \in \mathcal{V}_{h_i}, \ 2 \le h_i \le \kappa, \ \alpha_{ij} \in \mathbb{C}.
\end{equation}

Hence, the matrix $W_{f\!f}$ in \eqref{loi} can be transferred to an identity matrix $I_{n\!-\!m}$ by elementary row operations of the matrix $W_{f}$. Since the elementary row operations will not change the rank of the matrix $W_{f}$, we have
\begin{equation}\label{wdf}
\text{rank}(W_f)= \text{rank}(I_{n\!-\!m})=n-m.    
\end{equation}

Since $W_f \in \mathbb{C}^{(n\!-\!m)\times n}$ and $\text{rank}(W_f)=n\!-\!m$, the matrix $W_f$ is full row rank, i.e., the matrix $W_{f\!f}$ in \eqref{loi} is invertible. Hence, $p_f(t)$ in \eqref{loi} can be uniquely expressed by the positions of the leaders $p_l(t)$ and
the dynamic network $(\mathcal{G}, p
(t))$ is localizable. \ \ \ \ \ \ \ \ \ \ \ \ \ \ \  \ \ \ \ \  \ \ \ \ \ \ \ \ \ \ 
 \qed

\begin{figure}[t]
\centering
\includegraphics[width=0.9\linewidth]{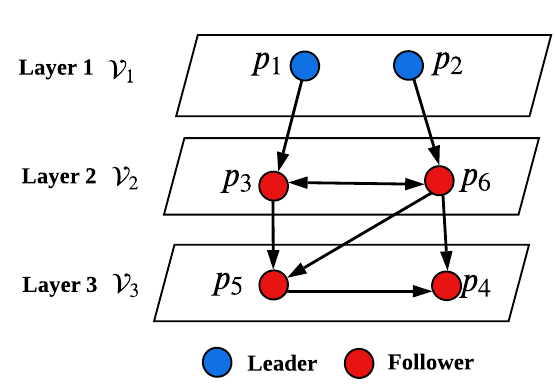}
\caption{Example of calculating the complex matrix $W_f$.}
\label{nece1}
\end{figure}

A simple example of calculating the complex matrix $W_f$ \eqref{g1} of a dynamic network $(\mathcal{G}, p(t))$ with a $\kappa$-layer graph is given in Fig. \ref{nece1}.  
For the followers $p_3, p_4, p_5, p_6$, we have
\begin{equation}\label{ex1}
\begin{array}{ll}
     &  w_{33}p_3 = w_{31}p_1 + w_{36}p_6, \ w_{44}p_4 = w_{46}p_6+w_{45}p_5, \\
     &  w_{55}p_5 = w_{53}p_3 + w_{56}p_6, \  w_{66}p_6 = w_{62}p_2 + w_{63}p_3. 
\end{array}
\end{equation}

Then, the complex matrix $W_f$ and configuration $p$ in \eqref{g1} are
\begin{equation}\label{ab}
\begin{array}{ll}
     &  p = (p_1^T,p_2^T,p_3^T,p_4^T,p_5^T,p_6^T)^T, \\
     &     W_f = \left[\begin{array}{cccccc}
    -w_{31} & 0 & \ w_{33} & 0 & 0 & -w_{36} \\
   0 & 0 & 0 & w_{44} & -w_{45} & -w_{46} \\
      0 & 0 & -w_{53} & 0 & \ w_{55} & -w_{56} \\
      0 & -w_{62} & -w_{63} & 0 & 0 & \ \ w_{66} \\
    \end{array}\right].
\end{array}
\end{equation}

From \eqref{ab}, we can know that the matrix $W_{f\!f}$ in \eqref{loi} is
\begin{equation}\label{ab1}
 W_{f\!f} = \left[\begin{array}{cccc}
     \ w_{33} & 0 & 0 & -w_{36} \\
 0 & w_{44} & -w_{45} & -w_{46} \\
       -w_{53} & 0 & \ w_{55} & -w_{56} \\
       -w_{63} & 0 & 0 & \ \ w_{66} \\
    \end{array}\right].
\end{equation}

The determinant of matrix $W_{f\!f}$ is
\begin{equation}
    \text{det}(W_{f\!f}) = w_{44}w_{55}(w_{33}w_{66}-w_{36}w_{63}).
\end{equation}

Under Assumption \ref{asu11}, we can know from Lemma \ref{le1} that the parameters $w_{33},w_{44},w_{55},w_{66} \neq 0$ and $w_{31},w_{36},w_{62},w_{63},$ $w_{46},w_{45},w_{53},w_{56} \neq 0$. 
If $\text{det}(W_{f\!f})=0$, we have $w_{33}w_{66}-w_{36}w_{63}=0$. From \eqref{ex1}, we obtain
\begin{equation}\label{tr}
    \frac{w_{31}}{w_{33}}(p_1 \!-\! p_2) = 0 \rightarrow p_1=p_2.
\end{equation}

Equation \eqref{tr} contradicts Assumption \ref{asu11} that $p_1 \! \neq \! p_2$. Hence, $\text{det}(W_{f\!f}) \neq 0$, i.e., the matrix $W_{f\!f}$ is invertible and the dynamic network $(\mathcal{G}, p(t))$ with a $\kappa$-layer graph in Fig. \ref{nece1} is localizable.

\begin{remark}
Compared with existing real-Laplacian-based or complex-Laplacian-based approaches in static networks, a remarkable advantage of the proposed method is that it is applicable to dynamic networks. 
\end{remark}

\section{Distributed Localization with Local Relative Positions}\label{protocol}

\begin{assumption}\label{as3}
The leaders have access to their positions. Each follower $i$ is equipped with a compass and has access to its velocity $v_i$.
\end{assumption}

\begin{remark}
The Assumption \ref{as3} that the leaders have access to their positions and the followers have access to their velocities can be removed by designing appropriate tracking controllers of the agents shown in Section \ref{formation}.  
\end{remark}

\begin{remark}\label{re5}
The local relative bearings are obtained by vision technology \cite{tron2016distributed}, where the unknown orientation difference $\theta_i(t)$ of agent $i$ is usually time-varying in order for keeping its neighbors within its field of view. Although the compass helps each follower know its moving orientation with respect to the north, the time-varying orientation difference $\theta_i(t)$ between $\Sigma_i$ and $\Sigma_g$ shown in Fig. \ref{sign}(b) is still unknown as the time-varying orientation difference $\theta_i(t)$ cannot be obtained by a compass. 
\end{remark}

\begin{remark}
Although compass may help the agents align the orientation between $\Sigma_i$ and $\Sigma_g$, it restricts the moving area and moving direction of the agents. Thus, it is more practical to consider the proposed case that different agent have different unknown time-varying orientation difference $\theta_i(t)$ in multi-agent systems.
\end{remark}

The distributed localization protocol is designed as
\begin{equation}\label{dis1}
\dot  {\hat p}_f =  - L_{f\!f} \hat p_f - L_{fl}p_l +  v_f,
\end{equation}
where $v_f=[v^H_{m\!+\!1} , \cdots, v^H_{n}]^H$ is the velocities of the followers and
\begin{equation}\label{dsl}
 L_{f\!f} =  W_{f\!f}^H W_{f\!f}, \   L_{fl} =  W_{f\!f}^H W_{fl}.
\end{equation}

The matrices $W_{f\!f}, W_{fl}$ given in \eqref{loi}
contain the complex weights among the agents, which
can be obtained by local relative positions over 
a $\kappa$-layer graph. 
To implement distributed localization protocol \eqref{dis1}, we need to use an augmented graph $ \mathcal{\bar G}$ defined as

\begin{defn}\label{aug}
Given a $\kappa$-layer graph $\mathcal{G}=(\mathcal{V},\mathcal{E})$, denote by $\mathcal{\bar G}=(\mathcal{V},\mathcal{\bar E})$ an augmented graph, where $\mathcal{\bar E} = \mathcal{E}  \cup  \{ (j,i): (i,j) \in \mathcal{E}, i,j \in \mathcal{V}_f \}$. 
\end{defn}

To obtain an augmented graph $\mathcal{\bar G}=(\mathcal{V},\mathcal{\bar E})$, we only need to make the directed edges among the followers of a $\kappa$-layer graph be undirected edges. Based on \eqref{dis1}, we can obtain the distributed localization protocol of each follower $i \in \mathcal{V}_f$, i.e.,
\begin{equation}\label{dis}
 \dot {\hat p}_i =  v_i + \delta_{ijk} +  \hspace{-0.1cm} \sum\limits_{i,s \in \mathcal{N}_h, h \in \mathcal
     {V}_f} \hspace{-0.5cm} \delta_{his}, 
\end{equation}
where $(h, i), (h,s) \in  \mathcal{E}$ and
\begin{equation}\label{disj}
    \begin{array}{ll}
         &  \delta_{ijk} = w_{ii}^Hw_{ij}(\hat p_j \!-\! \hat p_i)+w_{ii}^Hw_{ik}(\hat p_k \!-\! \hat p_i), \ j,k \in \mathcal{N}_i, \\
         & \delta_{his} = w_{hi}^Hw_{hi}(\hat p_h \!-\! \hat p_i) \!+\! w_{hi}^Hw_{hs}(\hat p_h \!-\! \hat p_s).
    \end{array}
\end{equation}

\begin{remark}
The localization protocol \eqref{dis} is implemented in a distributed manner: (1) The complex weights $w_{ii},w_{ij},w_{ik}$ in $\delta_{ijk}$ are calculated by its measured local relative positions through the directed edges $(i,j),(i,k) \in \mathcal{E}$, while the estimates of $\hat p_j, \hat p_k$ in $\delta_{ijk}$ are obtained by communication through the same directed edges $(i,j),(i,k) \in \mathcal{E}$. Thus, the information $\delta_{ijk}$ is calculated by each follower $i$ based on its measured local relative positions along with communication with its neighbors $j,k \in \mathcal{N}_i$; and (2) If follower $i$ is a neighbor of follower $h$ in $\mathcal{G}$, i.e., $(h,i) \in \mathcal{E}$, 
the complex weights $w_{hi},w_{hs}$ and estimates $\hat p_h,\hat p_s$ in $\delta_{his}$ are obtained by communicating with follower $h$ through the augmented edge $(i,h) \in \mathcal{\bar E}$.
\end{remark}

Define $e_f = \hat p_f - p_f$ as the position estimation error of the followers. Since $L_{f\!f}p_f+L_{fl}p_l= \mathbf{0}$,
it yields from \eqref{dis1} that
\begin{equation}\label{e1}
    \dot { e}_f = - L_{f\!f}  e_f.
\end{equation}

\begin{theorem}\label{theo2}
(\textbf{Stability Analysis}) 
Under Assumptions \ref{asu1}-\ref{as3}, given the initial position estimates $\hat p_i(0), i \in \mathcal{V}_f$ of the followers,  the distributed localization protocol \eqref{dis} can drive the position estimate of each follower 
$\hat p_i(t)$ to $p_i$ 
if the dynamic network is localizable.

\end{theorem}

\textbf{Proof.}
From Theorem \ref{alge}, we can know that the complex matrix $W_{f\!f}$ is invertible if the dynamic network is localizabe. The graph in Definition \ref{aug} can be used to guarantee that the dynamic network is localizabe at any time instant. 
Since $L_{f\!f}= W_{f\!f}^H W_{f\!f}$, the complex matrix $L_{f\!f}$ is a Hermitian matrix and positive definite. 
Since the matrix $L_{f\!f}$ is a complex matrix, the Lyapunov stability analysis method in real space cannot be used directly.  Note that the position estimation error $e_f \!=\! \hat p_f \!-\! p_f$ can also be rewritten as
\begin{equation}\label{f1}
e_f \!=\! e_{f_x} \!+\! e_{f_y}\iota, \ e_{f_x} \!=\! \hat x_{f} \!-\! x_f, \ e_{f_y} \!=\! \hat y_f \!-\! y_{f},
\end{equation}
where $e_{f_x}, e_{f_y} \in \mathbb{R}^{n\!-\!m}$ are the position estimation errors of the followers in real-axis and imaginary-axis, respectively. Let 
\begin{equation}\label{f2}
    L_{f\!f} = L_x + L_y \iota,
\end{equation}
where $L_{x}, L_{y} \in \mathbb{R}^{(n\!-\!m)\times(n\!-\!m)}$ are the real part and imaginary part of the complex matrix $L_{f\!f}$. The conjugate transpose of $L_{f\!f}$ is
\begin{equation}
    L^H_{f\!f} = L^T_x - L^T_y \iota.
\end{equation}

Since $L_{f\!f}$ is a Hermitian matrix, i.e., $L^H_{f\!f} = L_{f\!f}$, we can know that $L_{x}$ is a real symmetric matrix and $L_{y}$ is a real skew-symmetric matrix, i.e.,
\begin{equation}\label{f3}
   L_{x}^T = {L_{x}}, \  L_{y}^T + {L_{y}} = \mathbf{0}.
\end{equation}

Since $L_{f\!f}$ is positive definite, we have
\begin{equation}\label{f4}
\begin{array}{ll}
     &  e_f^HL_{f\!f}e_f =  e_{f_x}^TL_xe_{f_x} + e_{f_y}^TL_ye_{f_x} \\
     & \ \ \ \ \ \ \ \ \ \ \ \ \ \ \ \ \ \!- e_{f_x}^TL_ye_{f_y} \!+\! e_{f_y}^TL_xe_{f_y} >0,    \ \text{if} \ e_f \neq 0.
\end{array} 
\end{equation}

Based on \eqref{f1} and \eqref{f2},  \eqref{e1} is equivalent to
\begin{equation}\label{e2}
 \left\{ \! \! \begin{array}{lll} 
     \dot e_{f_x} \!=\! -L_{x} e_{f_x} \!+\! L_{y}e_{f_y},  \\
     \dot e_{f_y} \!=\! -L_{x} e_{f_y} \!-\! L_{y}e_{f_x}.
    \end{array}\right.
\end{equation}

Note that $e_f \neq 0$ if $e_{f_x}, e_{f_y} \neq 0$.
Consider a Lyapunov function $V_1 = \frac{1}{2}(e_{f_x}^T e_{f_x} + e_{f_y}^Te_{f_y})$, we have
\begin{equation}
\begin{array}{ll}
\dot V_1 &= e_{f_x}^T \dot e_{f_x} + e_{f_y}^T \dot e_{f_y} \\
& = e_{f_x}^T(-L_xe_{f_x} + L_ye_{f_y}) + e_{f_y}^T(-L_xe_{f_y} - L_ye_{f_x}) \\
& = -(e_{f_x}^TL_xe_{f_x} \!-\! e_{f_x}^TL_ye_{f_y}  \!+\! e_{f_y}^TL_ye_{f_x} \!+\! e_{f_y}^TL_xe_{f_y})  \\
& = -e_f^HL_{f\!f}e_f < 0, \ \text{if} \ e_{f_x}, e_{f_y} \neq 0.
\end{array}
\end{equation}

Hence, the position estimation error $e_f$ of the followers converges to zero. \ \ \ \ \ \ \ \ \ \ \ \ \ \ \ \ \ \ \ \  \ \ \ \ \ \ \ \ \ \ \ \ \ \ \ \ \ \ \ \ \ \ \ \ \ \ \ \ 
 \qed

\section{Distributed Localization with Distance and Sign of Direction Measurements}
\label{exte}

\subsection{Sign of Direction}

\begin{figure}[t]
\centering
\includegraphics[width=1\linewidth]{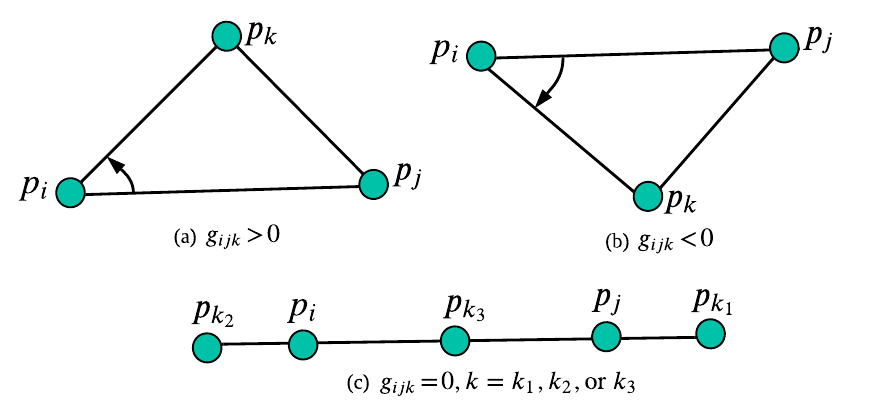}
\caption{Sign of direction measurement $g_{ijk}$.}
\label{sign1}
\end{figure}

The sign of direction measurement is 
denoted by
\begin{equation}\label{dsign}
 g_{ijk}\!=\! \text{sign}(e^{i}_{ij} \times e^{i}_{ik}),
\end{equation} 
where $\text{sign}(\cdot)$ is the signum function defined component-wise.  $g_{ijk}$ is the sign of direction among the agents $i,j,k$ shown in Fig. \ref{sign1}, i.e.,
\begin{equation}
\begin{array}{ll}
    &  \left\{ \! \begin{array}{lll} 
       g_{ijk}>0, \ p_i, p_j, p_k \ \text{are counterclockwise located,}  \\ 
       g_{ijk}<0, \ p_i, p_j, p_k \ \text{are clockwise located,} \\
       g_{ijk} = 0, \  p_i, p_j, p_k \ \text{are colinear.}
    \end{array}\right.  
\end{array} 
\end{equation}

\begin{remark}
$p_i, p_j, p_k$ are  counterclockwise located, i.e., it
travels in a counterclockwise direction if it goes in order from $p_i$ to $p_j$, then to $p_k$.
\end{remark}

The sign of direction $g_{ijk}$ is a kind of one bit local relative measurement, which can be extracted from a single image by vision technology \cite{tron2016distributed}.
The use of one bit relative measurement in multi-agent systems has been previously studied in  \citep{chen2011finite,guo2013consensus}.  The sign of direction is a rough local relative measurement, which is easier to obtain than those accurate local relative measurements such as local relative bearing and local relative position. Next, we will show how to calculate the complex weights $w_{ij}, w_{ik}$ in \eqref{complexc} by the distance and sign of direction measurements. 

\subsection{Distance and Sign of Direction based Complex Constraint}

\begin{defn}
Two configurations ${p}=(p^H_{1}, \cdots, p^H_{n})^H$ and ${q}=(q_1^H, \cdots, q^H_{n})^H$ are congruent if they have the same shape and size, i.e., 
$\|p_i\!-\!p_j\|_2=\| q_i\!-\!q_j\|_2$ for any two agents $i,j \in \mathcal{V}$.
\end{defn}

Let $\check p = (p_i^H, p^H_{j}, p^H_{k})^H$ be the configuration of agent $i$ and its two neighbors $j, k$. We can calculate its congruent configuration $\check{q}=(q_i^H, q^H_{j}, q^H_{k})^H$ by the distance measurements via the following steps:

(1) Set $q_i = 0$ and $q_j = d_{ij}$, where $d_{ij}=\|p_i \!-\! p_j\|_2$ is the distance between agent $p_i$ and agent $p_j$. Let 
\begin{equation}\label{qk}
q_k = x_{q_k} + y_{q_k} \iota,
\end{equation}
where $x_{q_k}, y_{q_k} \in \mathbb{R}$ are to be determined.
If $d_{ij}\!+\!d_{jk}\!=\!d_{ik}$, $d_{ij}\!+\!d_{ik}\!=\!d_{jk}$, or $d_{ik}\!+\!d_{kj}\!=\!d_{ij}$, agents $p_i, p_j, p_k$ are colinear shown in Fig. \ref{sign1}(c). Then, $q_k$ can be calculated as
\begin{equation}\label{xq}
 \begin{array}{ll}
    &  q_k = \left\{ \! \begin{array}{lll} 
      \ \ d_{ik},  \ \  d_{ij}\!+\!d_{jk}\!=\!d_{ik} \ \text{or} \ d_{ik}\!+\!d_{kj}\!=\!d_{ij}, \\
         - d_{ik}, \ \ d_{ij}\!+\!d_{ik}\!=\!d_{jk}.
\end{array}\right.  
\end{array} 
\end{equation}

(2) If $d_{ij}\!+\!d_{jk}\!\neq\!d_{ik}$, $d_{ij}\!+\!d_{ik}\!\neq\!d_{jk}$, and $d_{ik}\!+\!d_{kj}\!\neq\!d_{ij}$, agents $p_i, p_j, p_k$ are non-colinear shown in Fig. \ref{sign1}(a)-(b).  For $q_k$ in \eqref{qk}, we have
\begin{equation}\label{nonco}
    \begin{array}{ll}
         & x^2_{q_k}+y^2_{q_k}=d_{ik}^2, \\
         & (x_{q_k} \!-\! d_{ij})^2+y^2_{q_k}=d_{jk}^2.
    \end{array}
\end{equation}

From \eqref{nonco}, we obtain 
\begin{equation}\label{xq1}
\begin{array}{ll}
& x_{q_k} = \frac{d_{ij}^2+d_{ik}^2-d_{jk}^2}{2d_{ij}}, \\
& y_{q_k} = \pm \sqrt{d_{ik}^2 - x_{q_k}^2}.
\end{array} 
\end{equation}

Hence, for the configuration $\check p = (p_i^H, p^H_{j}, p^H_{k})^H$, we can calculate its congruent configuration $\check{q}=(q_i^H, q^H_{j}, q^H_{k})^H$ by the distance measurements through \eqref{xq} and \eqref{xq1}. Based on the sign of direction $g_{ijk}$,
\eqref{xq1} can be rewritten as
\begin{equation}\label{xq2}
\begin{array}{ll}
& x_{q_k} = \frac{d_{ij}^2+d_{ik}^2-d_{jk}^2}{2d_{ij}}, \\
&  y_{q_k} = \left\{ \! \begin{array}{lll} 
      \ \ \sqrt{d_{ik}^2 - x_{q_k}^2},  \ \  g_{ijk} >0, \\
         - \sqrt{d_{ik}^2 - x_{q_k}^2}, \ \ g_{ijk} < 0.
\end{array}\right.  
\end{array} 
\end{equation}

The sign of direction $g_{ijk}$ in \eqref{xq2} is used to distinguish the reflection of agents $p_i,p_j,p_k$ shown in Fig. \ref{sign1}(a)-(b). Thus, based on distance and sign of direction measurements,
we can find a congruent configuration $\check{q}=(q_i^H, q^H_{j}, q^H_{k})^H$ through \eqref{xq} and \eqref{xq2}, whose relationship to $\check p = (p_i^H, p^H_{j}, p^H_{k})^H$ is
\begin{equation}
 \check{q} = \beta +  \check p \cdot \text{exp}(\theta\iota),  
\end{equation}
where the parameters $\beta \! \in \! \mathbb{C}$ and $\theta \! \in \! [0, 2\pi]$ are unknown. From Section \ref{inva}, we can know that the congruent configuration $\check{q}=(q_i^H, q^H_{j}, q^H_{k})^H$ has the same complex weights $w_{ij}, w_{ik}$ as the configuration $\check p = (p_i^H, p^H_{j}, p^H_{k})^H$. Hence, the complex weights $w_{ij}, w_{ik}$ in \eqref{complexc} can be obtained by
\begin{equation}\label{weight1}
  w_{ij} = \frac{(q_j-q_i)^H}{d_{ij}^2}, \ w_{ik} = -\frac{(q_k-q_i)^H}{d_{ik}^2}.   
\end{equation}

After obtaining the complex weights among the agents by distance and sign of direction measurements, the positions of the followers in a dynamic network can also be estimated by the proposed distributed localization protocol \eqref{dis}.

\section{Integrated Distributed Localization and Formation Control}\label{formation}

Note that Assumption \ref{asu11} and Assumption \ref{as3} can be removed if we 
combine the distributed localization and formation control.

\subsection{Unknown Global Positions of the Leaders}

If the positions of all leaders in the global coordinate frame $\Sigma_g$ are unknown, we can build a virtual global coordinate frame $\digamma$ centered at the first leader, i.e., $p_1^{\digamma} = 0$, where the orientation difference between $\Sigma_g$ and $\digamma$ is set as zero. 
We consider the single-integrator kinematic model of each agent $i$ in $\digamma$, i.e.,
\begin{equation}
    \dot p_i^{\digamma} = v_i^{\digamma}, 
\end{equation}
where $v_i^{\digamma} \in \mathbb{C}$ is the velocity of agent $i$ in $\digamma$. Then, the control objectives of the leaders and followers  become
\begin{equation}\label{acon1}
\begin{array}{ll}
    &  \left\{ \! \begin{array}{lll} 
       \lim\limits_{t \rightarrow \infty} p_l^{\digamma}(t) = p^*_l(t), \\
       \lim\limits_{t \rightarrow \infty}\hat p_f(t) = p_f^{\digamma}(t), \\
        \lim\limits_{t \rightarrow \infty} p_f^{\digamma}(t) = p^*_f(t), 
    \end{array}\right.  
\end{array} 
\end{equation}
where $p^*_l \!=\! [(p^*_1)^H, \cdots, {(p^*_{m})}^H]^H$ and $p^*_f \!=\! [{(p^*_{m\!+\!1})}^H, \cdots, $ ${(p^*_{n})}^H]^H$ are the desired positions of the leaders and followers in $\digamma$, respectively. Let $p^*  \!=\! [(p^*_l)^H, {(p^*_{f})}^H]^H$ be the configuration of desired formation in  $\digamma$.

\subsection{Integrated Distributed Localization and Formation Control}

\begin{assumption}\label{asu2}
Each agent is equipped with a compass.
The positions of the rest leaders $p^{\digamma}_2, \cdots, p^{\digamma}_m$ in the virtual global coordinate frame $\digamma$ are known.  Each agent $i$ has access to its desired
formation parameters $p^*_i(t)$ and $\dot p_i^*(t)$.
\end{assumption}

\begin{remark}
For the case that only distance and sign of direction measurements are available in Section \ref{exte}, agent $p_i$ can  not be localized by a agent $p_j$
even if the time-varying orientation difference $\theta_i(t)$ is known since agent $p_i$ must have at least two neighbors to achieve self-localization. 
\end{remark}

\begin{remark}
If the first leader has access to the rest leaders, the positions of the rest leaders $p^{\digamma}_2, \cdots, p^{\digamma}_m$ in the virtual global coordinate frame $\digamma$ can be calculated 
based on the relative position measurements in $\digamma$.
\end{remark}

The integrated distributed localization and formation control protocol is designed as
\begin{equation}\label{int1}
\begin{array}{ll}
     & \  v_i^{\digamma} \!=\! -a(p_i^{\digamma}-p_i^*) + \dot p^*_i, \ i \in \mathcal{V}_l, \\
     &  \left\{ \! \! \begin{array}{lll}
   v_i^{\digamma} \!=\! -a(\hat p_i-p_i^*) \!+\! \dot p^*_i,  \\
    \dot {\hat p}_i \ \!=\! -2a(\hat p_i \!-\! p_i^*) \!+\! \dot p^*_i \!+\! \delta_{ijk}  \\
    \ \ \ \ \ \ \ \  \!+\!    \hspace{-0.1cm} \sum\limits_{i,s \in \mathcal{N}_{h}, h \in \mathcal
     {V}_f} \hspace{-0.5cm} \delta_{his}, 
    \end{array}\right. \ i \in \mathcal{V}_f,
\end{array}
\end{equation}
where ${\hat p}_i$ is the position estimate of follower $i$ in $\digamma$.
$a >0$ is a positive constant control gain, and $\delta_{ijk}, \delta_{his}$ are given in \eqref{disj}. From \eqref{int1}, we have
\begin{equation}\label{intee}
   \dot e_p = - L_p e_p, \ e_p = (\bar e_l^H, \bar e_f^H, e_f^H)^H,
\end{equation}
where $\bar e_l = p_l^{\digamma} - p_l^* \in \mathbb{C}^{m}$ and $\bar e_f = p_f^{\digamma} - p_f^* \in \mathbb{C}^{n\!-\!m}$ are the formation tracking errors of the leaders and followers, respectively. $e_f = \hat p_f - p_f^{\digamma} \in \mathbb{C}^{n\!-\!m}$ is the position estimation errors of the followers and
\begin{equation}\label{lp}
\begin{array}{ll}
     & L_p= \left[\begin{array}{lll}
   aI_m & \mathbf{0} &  \mathbf{0} \\
     \mathbf{0} & aI_{n\!-\!m} & aI_{n\!-\!m} \\
     \mathbf{0} & aI_{n\!-\!m} & aI_{n\!-\!m} \!+\! L_{f\!f}
    \end{array}\right].
\end{array}
\end{equation}

\begin{theorem}\label{intas}
(\textbf{Inter-agent Collision Avoidance and Stability Analysis}) 
Under Assumption \ref{asu1} and Assumption \ref{asu2},  given initial position estimates $\hat p_i(0), i \in \mathcal{V}_f$ of the followers,  the integrated distributed localization and formation control protocol \eqref{int1} can avoid  inter-agent collision and
achieve the control objective \eqref{acon1} if the dynamic network is localizable and the initial positions of the agents satisfy $\|p_i(0) \!-\! p_j(0)\|_2 \neq 0$ and $\|p_i^*(t) \!-\! p_j^*(t)\|_2 \!-\!  2\|e_p(0)\|_2 > 0$, $ \text{for any} \ i,j \in \mathcal{V}$.

\end{theorem}

\textbf{Proof.}
The condition $\|p_i^*(t)-p_j^*(t)\|_2 -  2\|e_p(0)\|_2 > 0$ is easy to be satisfied by designing the desired time-varying formation. For any $i,j \in \mathcal{V}$ and $t\ge0$, we obtain
\begin{equation}\label{eni}
    p_i- p_j = p_i^{\digamma}- p_j^{\digamma} =  (p_i^{\digamma} - p_i^*) + (p_j^*-p_j^{\digamma}) + (p_i^*-p_j^*). 
\end{equation}

Then, we have
\begin{equation}\label{trw}
\begin{array}{ll}
     &   \| p_i- p_j \|_2 \ge \|p_i^*-p_j^*\|_2 -  \|p_i^{\digamma} - p_i^*\|_2 - \|p_j^{\digamma}-p^*_j\|_2 \\
     & \ \ \ \ \ \ \ \ \ \ \ \ \  \ge \|p_i^*-p_j^*\|_2 -  2\|e_p\|_2.
\end{array}
\end{equation}

Note from \eqref{dsl} that 
the Hermitian matrix  $L_{f\!f} \ge 0$. Then, it yields from \eqref{lp} that the Hermitian matrix $L_p \ge 0$. 
Similar to the proof of Theorem \ref{theo2}, we can know from \eqref{intee} that 
\begin{equation}
    \|e_p(t)\|_2 \le \|e_p(0)\|_2.
\end{equation}

Then, \eqref{trw} becomes
\begin{equation}
 \| p_i(t)\!-\! p_j(t) \|_2 \!\ge\! \|p_i^*(t)\!-\!p_j^*(t)\|_2 \!-\!  2\|e_p(0)\|_2 \!>\! 0, t\ge0.   
\end{equation}

Hence, there will be no inter-collision among the agents at any time instant $t \ge 0$. From Theorem \ref{alge}, we can know that the complex matrix $W_{f\!f}$ is invertible if the dynamic network is localizabe. The graph $\mathcal{G}$ in Definition \ref{aug} can be used to guarantee that the dynamic network is localizabe at any time instant. Thus, 
$L_{f\!f}(t), L_p(t) > 0, t \ge 0$.  It yields from  \eqref{intee} that  $\lim\limits_{t \rightarrow \infty}e_p(t) =0$, i.e.,
the control objective \eqref{acon1} is achieved under the integrated distributed localization and formation control protocol \eqref{int1}. \ \ \ \ \ \ \ \ \ \ \ \ \ \ \ \ \ \ \ \  \ \ \ \ \ \ \ \ \ \ \ \ \ \ \ \ \ \ \ \ \ \ \ \ \ \ \ \ \ \ \ \ \ \ \ \ \ \ \ \ \ 
 \qed

\begin{remark}
Based on \eqref{lp} and \eqref{eni},
the condition $\|p_i^*(t) \!-\! p_j^*(t)\|_2 \!-\! 2\|e_p(0)\|_2 \! > \! 0$ can be relaxed as
\begin{equation}
  \|p_i^*(t)-p_j^*(t)\|_2 -   \lambda_i - \lambda_j > 0,
\end{equation}
where
\begin{equation}\label{kh}
\begin{array}{ll}
    &\lambda_i  \!=\!  \left\{ \! \begin{array}{lll} 
    \|p_i^{\digamma}(0) - p_i^*(0)\|_2, & i \in \mathcal{V}_l,  \\
   \| \left[\begin{array}{ll}
 \mathbf{0}   & \mathbf{0}  \\
     \mathbf{0} &  I_{2(n\!-\!m)} 
    \end{array}\right]e_p(0) \|_2,  &   i \in \mathcal{V}_f. \\
    \end{array}\right. 
\end{array} 
\end{equation}

From \eqref{intee}, we can calculate  $\|e_p(0)\|_2$ if the initial positions of the agents are known. For the case that the initial positions of the agents are not available, the agents are required to start integrated distributed localization and formation control in a bounded set such that the upper bounds $\|p_i^{\digamma}(0) \!-\! p_i^*(0)\|_2 \le \epsilon_i, i\in \mathcal{V}_l$ and $\| [\bar e^H_f(0), e_f^H(0)]^H\|_2 \le \epsilon_p$ are available. Then, \eqref{kh} is revised as
\begin{equation}\label{kh1}
\begin{array}{ll}
    &\lambda_i  \!=\!  \left\{ \! \begin{array}{lll} 
    \epsilon_i, & i \in \mathcal{V}_l,  \\
   \epsilon_p,  &   i \in \mathcal{V}_f. \\
    \end{array}\right. 
\end{array} 
\end{equation}

\end{remark}

\begin{figure*}[t]
\centering
\includegraphics[width=1\linewidth]{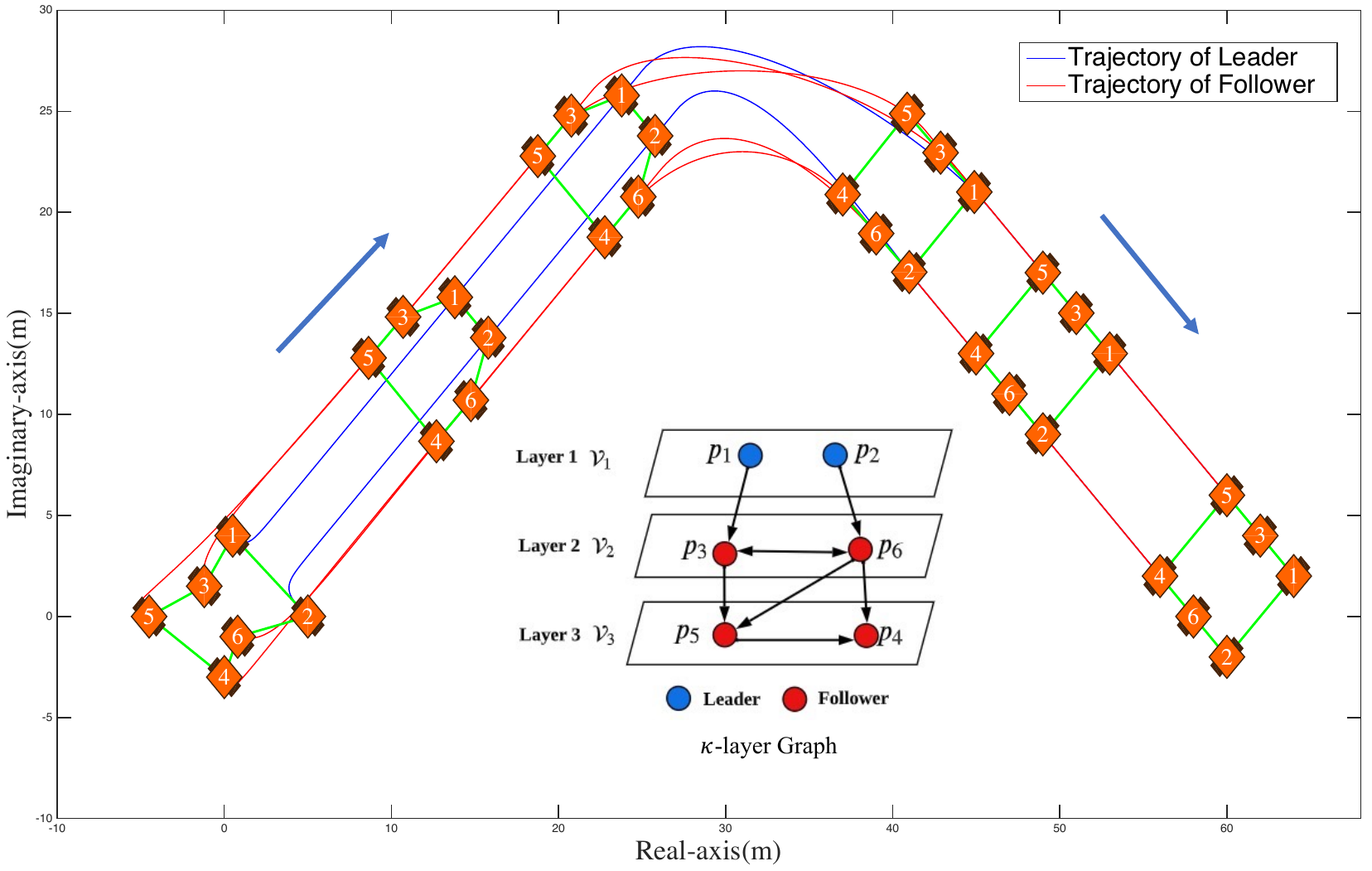}
\caption{Integrated distributed localization and formation control in dynamic network.}
\label{simu1}
\end{figure*}

\subsection{Distributed Parameter Estimator}

Let $p^*=[(p_1^*)^H, \cdots, (p_n^*)^H]^H$ be the time-varying desired formation in $\digamma$.
In some cases, only the leaders have access to the time-varying desired formation, i.e., $p^*(t)$ and $\dot p^*(t)$ in $\digamma$ are only available to the leaders. In a $\kappa$-layer graph $\mathcal{G}$ shown in Definition \ref{fn3}, for each follower, there exists at least one leader that has a direct path to that follower. Define $\mathcal{U}$ as a path of the $\kappa$-layer graph $\mathcal{G}$,  which includes a leader and $m_{\mathcal{U}}$ followers. We relabel $\mathcal{V}_{\mathcal{U}} = \{1,2, \cdots, m_{\mathcal{U}}\!+\!1 \}$ as the agent set in path $\mathcal{U}$, where $p_1$ is the leader and $p_2, \cdots, p_{m_{\mathcal{U}}\!+\!1}$ are the followers. A distributed parameter estimator is used for the followers to obtain their desired formation parameters, i.e., 
\begin{equation}\label{distpara}
  \dot {\psi}_i = \frac{1}{\eta_i} \sum\limits_{j=2}^{m_{\mathcal{U}}\!+\!1} b_{ij}[\dot {\psi}_j - \gamma ({\psi}_i-{\psi}_j)] + \frac{1}{\eta_i} b_{i1}[\dot {p}^* - \gamma ({\psi}_i-{p}^*)], 
\end{equation}
where ${\psi}_i(t)$ is the estimate of the formation parameter $p^*(t)$. $\gamma>0$ is a positive constant scalar and $\eta_i= \sum\limits_{j=1}^{m_{\mathcal{U}}\!+\!1} b_{ij}$. $b_{ij}\!=\!1$ if $(i, j) \in \mathcal{U}$ and $b_{ij}=0$ otherwise. 
 If follower $k$ is not on the path $\mathcal{U}$ but its neighbor $g \in \mathcal{N}_k$ is on the path $\mathcal{U}$, the parameter estimator of follower $k$ is given by
\begin{equation}
 \dot {\psi}_k = \dot {\psi}_g, \ {\psi}_k = {\psi}_g, \ g \in \mathcal{N}_k, g \in \mathcal{V}_{\mathcal{U}}.    
\end{equation}

Denote by $\mathcal{\bar V}_{\mathcal{U}} =  \mathcal{ V}_{\mathcal{U}}  \cup  \{ k: g \in \mathcal{N}_k, g \in \mathcal{ V}_{\mathcal{U}} \}$ an augmented agent set. 
From Theorem $3.8$ in \cite{ren2008distributed},
for each follower $i \in \mathcal{\bar V}_{\mathcal{U}}$, we have
\begin{equation}
\begin{array}{ll}
& \lim\limits_{t \rightarrow \infty}{\psi}_i(t) = p^*(t), \ \lim\limits_{t \rightarrow \infty} \dot {\psi}_i(t) = \dot p^*(t), \\
&   \|{\psi}_i(t) - p^*(t)\|_2 \le \sum\limits_{i=2}^{m_{\mathcal{U}}\!+\!1} \|{\psi}_i(0) - p^*(0)\|_2.
\end{array}
\end{equation}

From Theorem $3.8$ in \cite{ren2008distributed}, the parameter estimation error  $\|{\psi}_i(t) - p^*(t)\|_2$ converges to zero exponentially fast. In a $\kappa$-layer graph $\mathcal{G}$, we can assign all followers to $\varrho$ disjoint augmented agent set $\mathcal{\bar V}_{\mathcal{U}_1}, \cdots, \mathcal{\bar V}_{\mathcal{U}_{\varrho}}$, where each augmented agent set consists of a leader and $m_{\mathcal{U}_j}(j=1, \cdots, \varrho)$ followers. Let
\begin{equation}\label{zet}
    e_{\zeta}(0) = \max\limits_{j =1, \cdots, \varrho}  \sum\limits_{i=2}^{m_{\mathcal{U}_j}\!+\!1} \|{\psi}_i(0) - p^*(0)\|_2 .
\end{equation}

Note that ${\psi}_i = [(\tilde p_1^*)^H, \cdots, (\tilde p_n^*)^H]^H$ and ${\dot \psi}_i = [(\dot {\tilde p}_1^*)^H, \cdots, (\dot {\tilde p}_n^*)^H]^H$ are the estimated desired positions and velocities of the multi-agent system. Each follower $i$ can obtain its desired position $\tilde p_i^*$ and velocity $\dot {\tilde p}_i^*$ from ${\psi}_i$ and ${\dot \psi}_i$ \eqref{distpara}.
Then, Assumption \ref{asu2} is relaxed as

\begin{assumption}\label{asu3}
Each agent is equipped with a compass.
The positions of the rest leaders $p^{\digamma}_2, \cdots, p^{\digamma}_m$ in the virtual global coordinate frame $\digamma$ are known. Only the leaders have access to the
time-varying desired formation, i.e., $p^*(t)$ and $\dot p^*(t)$ in $\digamma$ are only available to the leaders.
\end{assumption}

\begin{figure}[t]
\centering
\includegraphics[width=1\linewidth]{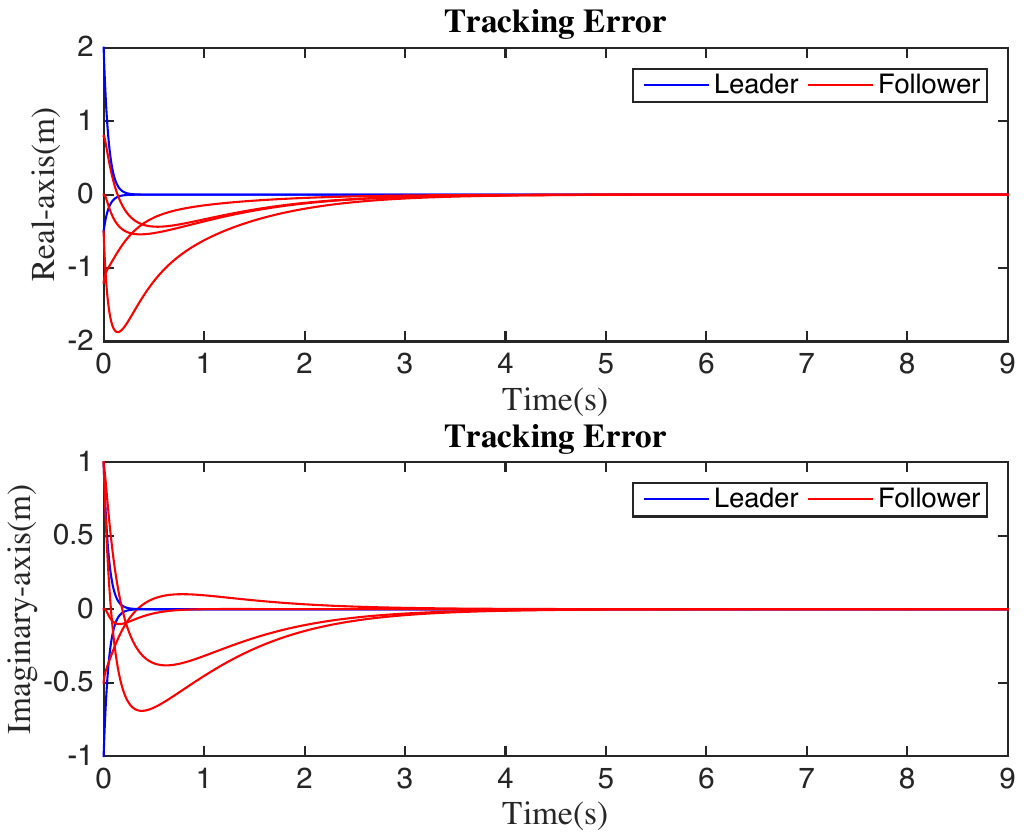}
\caption{Tracking errors of the agents.}
\label{simu2}
\end{figure}

The integrated distributed localization and formation control protocol in \eqref{int1} is revised as
\begin{equation}\label{int2}
\begin{array}{ll}
     & \  v^{\digamma}_i \!=\! -a(p^{\digamma}_i-p_i^*) + \dot p^*_i, \ i \in \mathcal{V}_l, \\
     &  \left\{ \! \! \begin{array}{lll}
   v^{\digamma}_i \!=\! -a(\hat p_i- \tilde p_i^*) + \dot {\tilde p}^*_i,  \\
    \dot {\hat p}_i = -2a(\hat p_i- \tilde p_i^*) \!+\! \dot {\tilde p}^*_i \!+\! \delta_{ijk} \\
    \ \ \ \ \ \ \ \  \!+\!    \hspace{-0.1cm} \sum\limits_{i,s \in \mathcal{N}_{h}, h \in \mathcal
     {V}_f} \hspace{-0.5cm} \delta_{his}, 
    \end{array}\right.  \ i \in \mathcal{V}_f.
\end{array} 
\end{equation}

The formation tracking errors of the followers $\bar e_f$ in $e_p$ \eqref{intee} is revised as $\bar e_f = p_f - \tilde p_f^*$.

\begin{corollary}\label{corro1}
Under Assumption \ref{asu1} and Assumption \ref{asu3},  given the initial position estimates $\hat p_i(0), i \in \mathcal{V}_f$ of the followers,  the integrated distributed localization and formation control protocol \eqref{int2} can avoid  inter-agent collision and
achieve the control objective \eqref{acon1} if the dynamic network is localizable and the initial positions of the agents satisfy $\|p_i(0) \!-\! p_j(0)\|_2 \neq 0$ and $\|p_i^*(t) \!-\! p_j^*(t)\|_2 \!-\!  2\|e_p(0)\|_2 \!-\! 2\|e_{\zeta}(0)\|_2 > 0$, $ \text{for any} \ i,j \in \mathcal{V}$. 
\end{corollary}

The proof of Corollary \ref{corro1} follows from that of Theorem \ref{intas}.

\begin{figure}[t]
\centering
\includegraphics[width=1\linewidth]{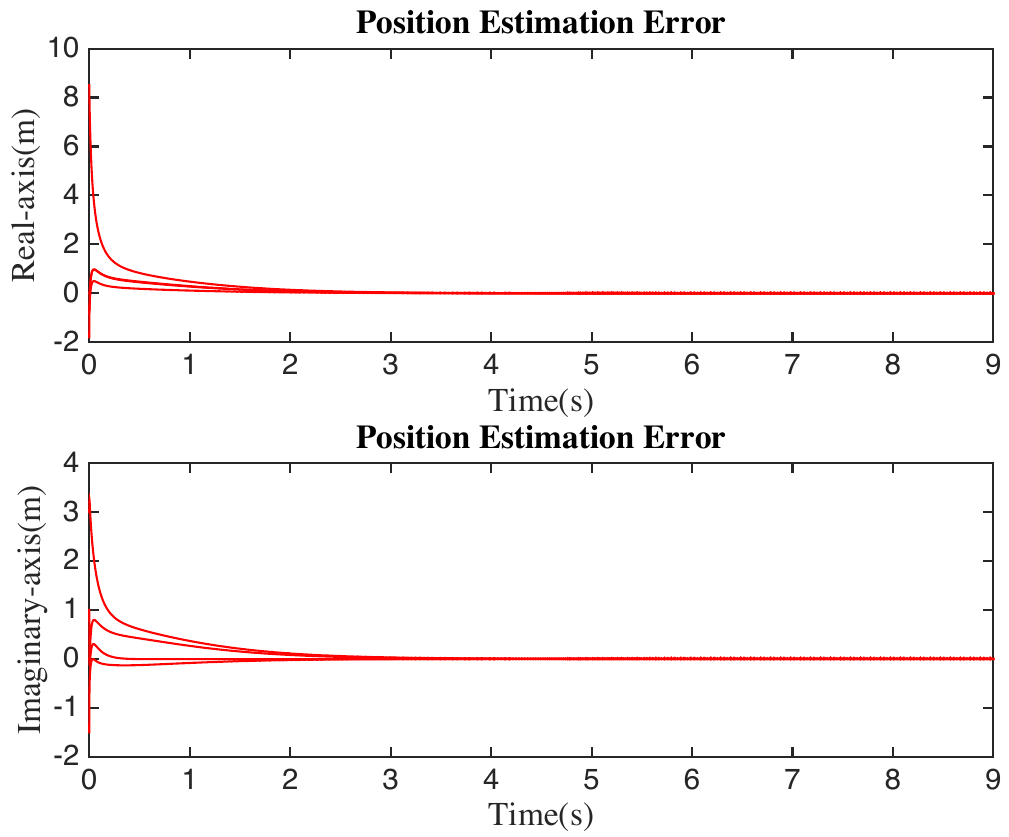}
\caption{Position estimation errors of the followers.}
\label{simu3}
\end{figure}

\section{Simulation}\label{simul}

An integrated distributed localization and formation control, consisting of two leaders $\mathcal{V}_l=\{ 1,2 \}$ and four followers $\mathcal{V}_f=\{ 3,4,5,6 \}$, is presented in this section to verify the proposed theoretical results. Each agent can obtain distance and sign of direction measurements in its unknown local coordinate frame. 
The designed $\kappa$-layer graph is given in Fig. \ref{nece1}.  The initial positions of the agents are given as
\begin{equation}
\begin{array}{ll}
     & p_1 = 1+3\iota, \ \ p_2 = 3+\iota, \ \ p_3 = -2+2\iota,  \\
     & p_4 = -4+\iota, \ \ p_5 = -4, \ \ p_6=2-2\iota.
\end{array}
\end{equation}

\begin{figure}[t]
\centering
\includegraphics[width=1\linewidth]{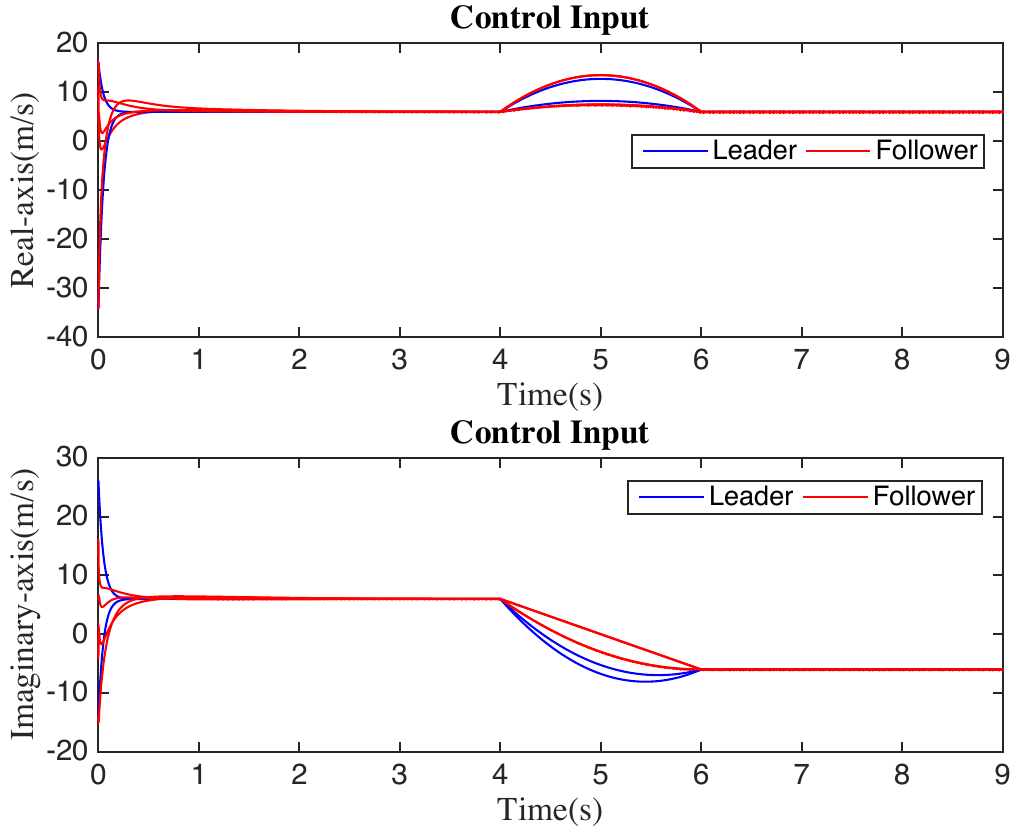}
\caption{Control inputs of the agents.}
\label{simu4}
\end{figure}

The corresponding complex matrix $W_f$ of Fig. \ref{nece1} is given in \eqref{ab}. Its submatrix $W_{f\!f}$ in \eqref{ab1} is invertible at any time instant $t$ if $\|p_i^*(t)-p_j^*(t)\|_2 -  2\|e_p(0)\|_2 > 0$ for any $i,j \in \mathcal{V}$, i.e., the dynamic network is localizable. 
It is shown in Fig. \ref{simu1} that the agents can form the desired formation and avoid inter-collision under the proposed integrated distributed localization and formation control protocol \eqref{int1}. The formation tracking errors  of the leaders and followers converge to zero shown in Fig. \ref{simu2}, while the position estimation errors of the followers also converge to zero  given in Fig. \ref{simu3}.  The control inputs of the agents are shown in Fig. \ref{simu4}. During the time interval $t \in [4, 6]$, all agents change their moving directions to make a turn.
Thus, the control inputs of all agents vary during the time interval $t \in [4, 6]$.

\section{Conclusion}\label{conc}

This work studies the distributed localization problem in dynamic networks, where only distance and local bearing or sign of direction measurements are available. Both algebraic condition and graph condition are given to guarantee a dynamic network to be localizable. Then, a distributed localization protocol is used to drive the position estimation errors to zero. The proposed method is extended to solve the problem of integrated distributed localization and formation control in dynamic networks, where the agents can avoid inter-collision. 

 \bibliographystyle{plain}

 \bibliography{autosam}

\balance

\end{document}